\documentclass[twocolumn,pre,aps,showpacs,amsmath,amssymb]{revtex4}
\usepackage{graphicx}

\begin{document}

\title{Strange Nonchaotic Bursting in the Quasiperiodically Forced
Hindmarsh-Rose Neuron}

\author{Woochang Lim}
\email{wclim@kangwon.ac.kr} \affiliation{Department of Physics,
Kangwon National University, Chunchon, Kangwon-Do 200-701, Korea}
\author{Sang-Yoon Kim}
\thanks{Corresponding Author}
\email{sykim@kangwon.ac.kr} \affiliation{Department of Physics,
Kangwon National University, Chunchon, Kangwon-Do 200-701, Korea}

\begin{abstract}
We study the transition from a silent state to a bursting state by varying the dc stimulus in the Hindmarsh-Rose neuron under quasiperiodic stimulation. For this quasiperiodically forced case, a new type of strange nonchaotic (SN) bursting state is found to occur between the silent state and the chaotic bursting state. This is in contrast to the periodically forced case where the silent state transforms directly to a chaotic bursting state. Using a rational approximation to the quasiperiodic forcing, the mechanism for the appearance of such an SN bursting state is investigated. Thus, a smooth torus (corresponding to a silent state) is found to transform to an SN bursting attractor through a phase-dependent subcritical period-doubling bifurcation. These SN bursting states, together with chaotic bursting states,
are characterized in terms of the interburst interval, the bursting length, and the number of spikes in each burst. Both bursting states are found to be aperiodic complex ones. Consequently, aperiodic complex burstings may result from two dynamically different states with strange geometry (one is chaotic  and the other one is nonchaotic). Thus, in addition to chaotic burstings, SN burstings may become a dynamical origin for complex physiological rhythms which are ubiquitous in organisms.
\end{abstract}


\maketitle

\section{Introduction}
\label{sec:INT}

To probe dynamical properties of a system, one often applies an
external stimulus to the system and study its response.
Particularly, periodic stimulation to biological oscillators has
attracted much attention in various systems such as the embryonic
chick heart-cell aggregates \cite{Chicken}, the squid giant
axon \cite{Squid,Squid2}, and the cortical pyramidal neurons \cite{Stoop}. Rich regular (such as phase locking and quasiperiodicity) and chaotic responses were found in these periodically forced systems \cite{Glass1,Glass2}. In contrast, quasiperiodically forced case has received little attention \cite{QO,QKim}, and hence further intensive investigation on dynamical responses of quasiperiodically forced biological oscillators is necessary.

Here, we are interested in neural bursting activity [alternating between a silent phase and an active (bursting) phase of repetitive spikings] \cite{Bursting1}. Cortical intrinsically bursting neurons, thalamocortical relay neurons, thalamic reticular neurons, and hippocampal pyramidal neurons are representative examples of bursting neurons \cite{Bursting}. We are particularly concerned about dynamical responses of bursting neurons subject to quasiperiodic stimulation. Strange nonchaotic (SN) attractors typically appear in quasiperiodically forced dynamical systems \cite{Greb,SNA,PD,Kim}. They exhibit some properties of regular as well as chaotic attractors. Like
regular attractors, their dynamics is nonchaotic in the sense that
they do not have a positive Lyapunov exponent; like usual chaotic
attractors, they have a geometrically strange (fractal) structure.
Hence, SN burstings are expected to occur in quasiperiodically
forced bursting neurons.

This paper is organized as follows. In Sec.~\ref{sec:IT}, we consider the Hindmarsh-Rose (HR) neuron model for bursting neurons which was originally  introduced to describe the time evolution of the membrane potential for the pond snail \cite{Bursting1,HR,HR1}, and investigate the transition from a silent state to a bursting state by varying the dc stimulus.
This work is in contrast to previous works on the effect of the quasiperiodic forcing on the self-oscillating neurons in the spiking state of self-sustained oscillations of the membrane potential \cite{QKim}. In the periodically forced case ({\it i.e.}, in the presence of only one ac stimulus source), an intermittent transition from a silent state (with subthreshold oscillations) to a chaotic bursting state occurs when the dc stimulus passes a threshold value. Effect of the quasiperiodic forcing on this intermittent route to chaotic bursting is investigated by adding another independent ac stimulus source. Thus, unlike the case of periodic stimulus, a new type of SN burstings are found to occur between the silent state and chaotic bursting state as intermediate ones. Using a rational approximation to the quasiperiodic forcing \cite{PD,Kim}, we investigate the mechanism for the appearance of such SN
burstings. Thus, a smooth torus, corresponding to a silent state,
is found to transform to an SN bursting attractor via a
phase-dependent subcritical period-doubling bifurcation. Together
with chaotic burstings, these SN burstings are characterized in
terms of the interburst interval, the bursting length, and the
number of spikes in each burst. Both the chaotic and SN bursting
states are found to be aperiodic complex ones. Such aperiodic
complexity comes from the strange geometry of both bursting states
with qualitatively different dynamics (one is chaotic and the
other one is nonchaotic). We note that complex physiological rhythms, which are central to life, are ubiquitous in organisms \cite{Glass2}. Hence, in addition to chaotic burstings, SN burstings may also serve as a dynamical origin of such complex bodily rhythms. Finally, a summary is given in Sec.~\ref{sec:SUM}.

\section{SN Burstings in the Quasiperiodically Forced HR Neuron}
\label{sec:IT}

We consider a representative HR bursting neuron model \cite{Bursting1,HR,HR1}
which is quasiperiodically forced at two incommensurate
frequencies $f_1$ and $f_2$:
\begin{subequations}
\begin{eqnarray}
\frac{dx}{dt} &=& y - a x^3 + b x^2 -z + I_{\rm ext}, \\
\frac{dy}{dt} &=& c - d x^2 - y, \\
\frac {dz} {dt} &=& r [ s (x-x_0) -z],
\end{eqnarray}
\label{eq:HR1}
\end{subequations}
where $I_{\rm ext} = I_{\rm dc} + A_1 \sin(2 \pi f_1 t) + A_2
\sin(2 \pi f_2 t)$, $a=1$, $b=3$, $c=1$, $d=5$, $s=1$, $r=0.001$,
and $x_0 = -1.6$. Here, $t$ is the time [measured in units of millisecond (ms)], $x$ is the membrane potential variable, $y$ is the recovery
variable, $z$ is the slow adaptation current, $I_{\rm dc}$ is a dc
stimulus, $A_1$ and $A_2$ are amplitudes of quasiperiodic forcing,
and $\omega (\equiv  f_2 / f_1)$ is irrational ($f_1$ and $f_2$:
measured in units of kHz).

To obtain the Poincar\'{e} map of Eq.~(\ref{eq:HR1}), we make a
normalization $f_1 t \rightarrow t$, and then Eq.~(\ref{eq:HR1})
can be reduced to the following differential equations:
\begin{subequations}
\begin{eqnarray}
\frac{dx}{dt} &=& F_1({\bf{x}},\theta) = {\frac {1} {f_1}}
(y - a x^3 + b x^2 -z + I_{\rm ext}), \\
\frac{dy}{dt} &=& F_2({\bf{x}},\theta) = {\frac {1} {f_1}} (c - d x^2 - y), \\
\frac {dz} {dt} &=& F_3({\bf{x}},\theta) = {\frac {r} {f_1}}  [ s
(x-x_0) -z], \\
\frac {d\theta} {dt} &=& \omega~~{\rm {(mod~1)}},
\end{eqnarray}
\label{eq:HR2}
\end{subequations}
where ${\bf x} = (x,y,z)$ and $I_{\rm ext}= I_{\rm dc} + A_1
\sin(2 \pi t) + A_2 \sin(2 \pi \theta)$. The phase space of the
quasiperiodically forced HR oscillator is five dimensional with
coordinates $x$, $y$, $z$, $\theta$, and $t$. Since the system is
periodic in $\theta$ and $t$, they are circular coordinates in the
phase space. Then, we consider the surface of section, the
$x$-$y$-$z$-$\theta$ hypersurface at $t=n$ ($n$: integer). The
phase-space trajectory intersects the surface of section in a
sequence of points. This sequence of points corresponds to a
mapping on the four-dimensional hypersurface. The map can be
computed by stroboscopically sampling the orbit points ${\bf v}_n$
$[\equiv ({\bf x}_n,\theta_n)]$ at the discrete time $n$
(corresponding to multiples of the first external driving period
$T_1$). We call the transformation ${\bf v}_n \rightarrow {\bf
v}_{n+1}$ the Poincar{\'e} map, and write ${\bf v}_{n+1} = P({\bf
v}_n)$.

Numerical integration of Eqs.~(\ref{eq:HR1}) and (\ref{eq:HR2}) is
done using the fourth-order Runge-Kutta method. Dynamical analysis
is performed in both the continuous-time system ({\it i.e.}, flow)
and the discrete-time system ({\it i.e.}, Poincar\'{e} map). For
example, the time series of the membrane potential $x(t)$, the
phase flow, the interburst interval, the bursting length, and the
average number of spikes per burst are obtained in the flow. On
the other hand, the Lyapunov exponent \cite{Lexp} and the phase
sensitivity exponent \cite{PD} of an attractor are calculated in
the Poincar\'{e} map. To obtain the Lyapunov exponent of an
attractor in the Poincar\'{e} map, we choose 20 random initial
points $\{ (x_i(0), y_i(0), z_i(0), \theta_i(0)); i=1,\dots,20 \}$
with uniform probability in the range of $x_i(0) \in (-2,2)$,
$y_i(0) \in (-16,0)$, $z_i(0) \in (0,0.4)$, and $\theta_i(0) \in
[0,1)$. For each initial point, we get the Lyapunov exponent
\cite{Lexp}, and choose the average value of the 20 Lyapunov
exponents. (The method of obtaining the phase sensitivity exponent
will be explained below.)

\begin{figure}
\includegraphics[width=\columnwidth]{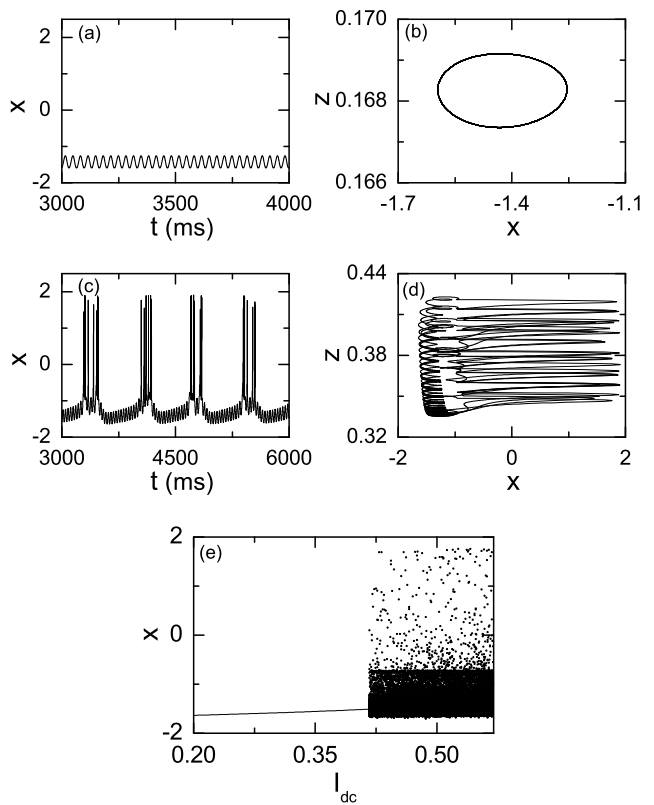}
\caption{Intermittent transition to a chaotic bursting state for
the case of periodic forcing with $A_1=0.5$ and $f_1=30$ Hz
($A_2=0$). (a) Time series of $x(t)$ and (b) projection of the
phase flow onto the $x-z$ plane for the silent state exhibiting
subthreshold oscillations for $I_{\rm dc}=0.3$. (c) Time series of
$x(t)$ and (d) projection of the phase flow onto the $x-z$ plane
for the chaotic bursting state for $I_{\rm dc}=0.5$. (e)
Bifurcation diagram ({\it i.e.}, plot of $x$ versus $I_{\rm dc}$)
in the Poincar\'{e} map. We obtain attractors by iterating the
Poincar\'{e} map at the 500 equally-spaced values of $I_{\rm dc}$
in the range of $I_{\rm dc} \in [0.2,0.57]$. For each chosen
$I_{\rm dc}$, we choose a random initial point $(x(0), y(0), z(0),
\theta(0))$ with uniform probability in the range of $x(0) \in
(-2,2)$, $y(0) \in (-16,0)$, $z(0) \in (0,0.4)$, and $\theta(0)
\in [0,1)$, and obtain the attractor through the 200-times
iterations of the Poincar\'{e} map after the transients of the
1000 Poincar\'{e} maps. \label{fig:CB}}
\end{figure}

Here, we set $\omega$ to be the reciprocal of the golden mean
[{\it i.e.}, $\omega = (\sqrt{5}-1)/2$], and numerically
investigate dynamical transition from a silent state to a bursting state
by varying $I_{dc}$ in the HR neuron under external stimulus. We first consider the case of periodic forcing ({\it i.e.}, $A_2=0$) for $A_1=0.5$ and $f_1=30$ Hz. Figures \ref{fig:CB}(a) and \ref{fig:CB}(b) show the time series of $x(t)$
and the projection of the phase flow onto the $x$-$z$ plane for
the silent state when $I_{\rm dc}=0.3$. We note that this silent
state with the largest Lyapunov exponent $\sigma_1 \simeq -0.133$
exhibits subthreshold oscillations. As $I_{\rm dc}$ passes a
threshold value of $I_{\rm dc}=0.416\,721$, a chaotic bursting
state appears. Bursting activity [alternating between a silent
phase and an active (bursting) phase of repetitive spikings] of the HR
neuron is shown in Fig.~\ref{fig:CB}(c) for $I_{\rm dc}=0.5$. This
kind of bursting occurs on a chaotic hedgehoglike attractor with
$\sigma_1 \simeq 0.406$ [the body (spines) of the hedgehoglike
attractor corresponds to the silent (bursting) phase], as shown in
Fig.~\ref{fig:CB}(d). This transition from a silent state to a
chaotic bursting state is investigated by varying $I_{\rm dc}$ in
the Poincar\'{e} map. Figure \ref{fig:CB}(e) shows the bifurcation
diagram ({\it i.e.}, plot of $x$ versus $I_{\rm dc}$). The solid
curve represents a stable fixed point corresponding to the silent
state. The stable fixed point loses its stability for $I_{\rm
dc}=0.416\,721$ via a subcritical Hopf bifurcation when a pair of
complex conjugate stability multipliers passes the unit circle in
the complex plane, and then a chaotic bursting attractor,
corresponding to a chaotic bursting state, appears.

\begin{figure}
\includegraphics[width=\columnwidth]{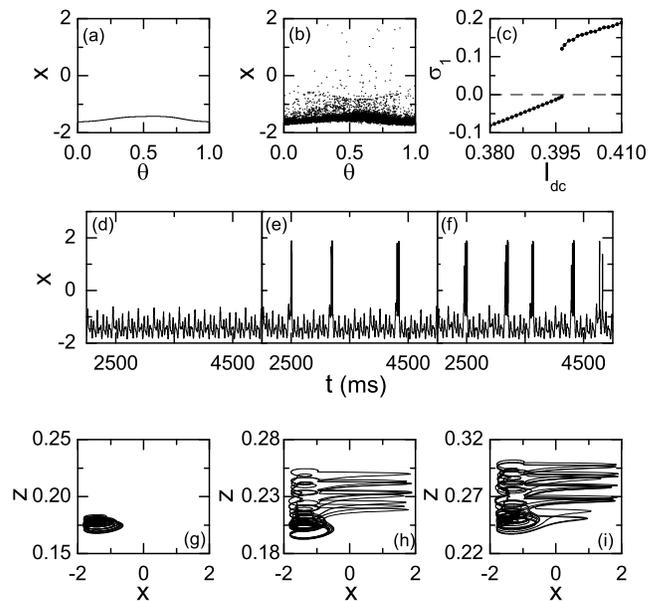}
\caption{Transition from a silent state to a bursting state for the
quasiperiodically forced case when $A_1=0.5$ and $f_1=30$ Hz. We
set $A_2=0.2$ in (a)-(c). Projections of attractors onto the
$\theta-x$ plane in the Poincar\'{e} map are shown for the (a)
silent and (b) chaotic bursting states when $I_{\rm dc} = 0.39$,
and 0.4, respectively. (c) Lyapunov-exponent diagram ({\it i.e.,}
plot of $\sigma_1$ versus $I_{\rm dc}$). For each randomly chosen
initial point, we get the Lyapunov exponent by following the
$10^4$ Poincar\'{e} maps after the transients of the $10^4$
Poincar\'{e} maps. We set $A_2=0.5$ for (d)-(i). Time series of
$x(t)$ of (d) the silent state for $I_{\rm dc} = 0.21$, (e) the SN
bursting state for $I_{\rm dc} = 0.24$, and (f) the chaotic
bursting state for $I_{\rm dc} = 0.29$. Projections of the phase
flows onto the $x-z$ plane for (g) the silent state for $I_{\rm
dc} = 0.21$, (h) the SN bursting state for $I_{\rm dc} = 0.24$,
and (i) the chaotic bursting state for $I_{\rm dc} = 0.29$.
\label{fig:SNB1}}
\end{figure}

From now on, we consider the quasiperiodically forced case for
$A_1=0.5$ and $f_1=30$ Hz. Each state is characterized by both the
largest (nontrivial) Lyapunov exponent $\sigma_1$, associated with
dynamics of the variable ${\bf x}$ [besides the (trivial) zero
exponent, related to the phase variable $\theta$ of the
quasiperiodic forcing] and the phase sensitivity exponent
$\delta$. The exponent $\delta$ measures the sensitivity of the
variable ${\bf x}$ with respect to the phase $\theta$ of the
quasiperiodic forcing and characterizes the strangeness of an
attractor \cite{PD}. A (regular) silent state has a negative
Lyapunov exponent ({\it i.e.}, $\sigma_1 <0$) and has no phase
sensitivity ({\it i.e.}, $\delta=0$). On the other hand, a chaotic
bursting state has a positive Lyapunov exponent $\sigma_1>0$. In
addition to them, a new type of SN bursting states that have
negative Lyapunov exponents ($\sigma_1<0$) and positive phase
sensitivity exponents $(\delta>0)$ appear. Due to their high phase
sensitivity, SN bursting states have a strange fractal phase space
structure. For small $A_2$, a direct transition from a silent
state to a chaotic bursting state occurs, as in the periodically
forced case of $A_2=0$. As an example, we consider the case of
$A_2=0.2$ where a transition to chaotic bursting occurs for
$I_{\rm dc}=0.3963$. Figures \ref{fig:SNB1}(a) and
\ref{fig:SNB1}(b) show a smooth torus with $\sigma_1 \simeq
-0.036$ (corresponding to a silent state) and a chaotic bursting
attractor with $\sigma_1 \simeq 0.154$ (corresponding to a chaotic
bursting state) for $I_{\rm dc}=0.39$ and 0.4, respectively.
Figure \ref{fig:SNB1}(c) shows the Lyapunov-exponent diagram ($\it
i.e.,$ plot of $\sigma_1$ vs. $I_{\rm dc}$). As $I_{\rm dc}$ is
increased to the transition point, $\sigma_1$ of the smooth torus
increases to zero, and then a chaotic bursting attractor with a
(finite) positive $\sigma_1$ appears ($\it i.e.,$ a finite jump
for the value of $\sigma_1$ seems to occur). However, for $A_2$
larger than a threshold $A_2^*$ $(\sim 0.4)$, SN bursting states
appear between the silent and chaotic bursting states. As an
example, we consider the case of $A_2=0.5$ and investigate
dynamical behaviors of the quasiperiodically forced HR neuron by
varying $I_{\rm dc}$. As $I_{\rm dc}$ passes a threshold $I^*_{\rm
dc}$ $(\simeq 0.2236)$, the silent state becomes unstable, and a
transition to an SN bursting state occurs. As $I_{\rm dc}$ is
further increased and passes another threshold value of $I_{\rm
dc} \simeq 0.2703$, the SN bursting state transforms to a chaotic
bursting state. Figures \ref{fig:SNB1}(d)-\ref{fig:SNB1}(f) show
the time series of the membrane potential $x(t)$ of a silent state
(exhibiting subthreshold oscillations), an SN bursting state, and
a chaotic bursting state for $I_{\rm dc}=$ 0.21, 0.24, and 0.29,
respectively. For these three cases, the phase flows of the
silent, SN bursting and chaotic bursting states are also given in
Figs.~\ref{fig:SNB1}(g)-\ref{fig:SNB1}(i), respectively.

\begin{figure}
\includegraphics[width=\columnwidth]{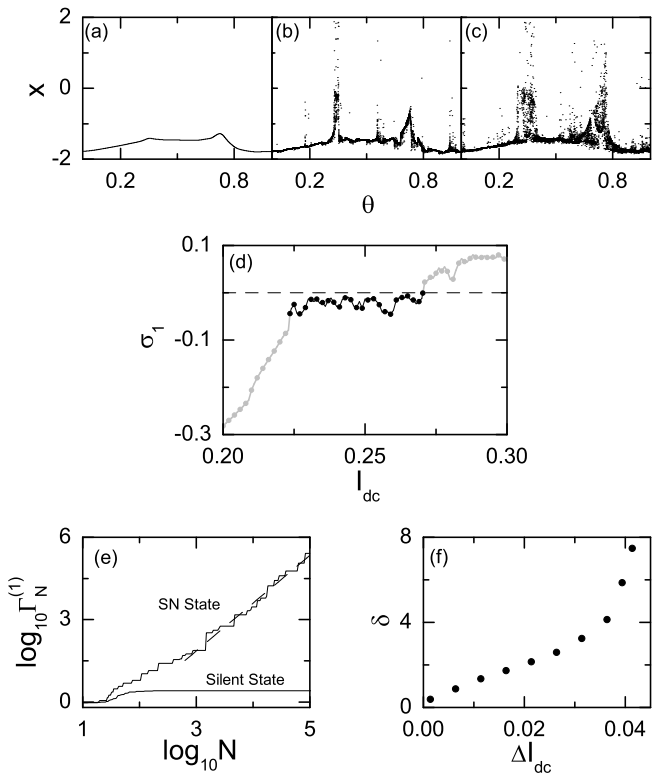}
\caption{Characterization of silent and bursting states for $A_1=0.5$,
 $A_2=0.5$, and and $f_1=30$ Hz. Projections of attractors onto the $\theta-x$ plane in the Poincar\'{e} map are shown for the (a) silent, (b) SN bursting, and (c) chaotic bursting states when $I_{\rm dc}$ = 0.21, 0.24, and 0.29,
respectively. (d) Lyapunov-exponent diagram ({\it i.e.,} plot of
$\sigma_1$ versus $I_{\rm dc}$); $\sigma_1$ for the SN bursting
attractor is shown in black. For each randomly chosen initial
point, we get the Lyapunov exponent by following the $10^4$
Poincar\'{e} maps after the transients of the $10^3$ Poincar\'{e}
maps. (e) Phase sensitivity functions $\Gamma^{(1)}_N$ are shown
for the silent and SN bursting attractors when $I_{\rm dc}$=0.21
and 0.24, respectively. For the case of SN bursting attractor, the
graph is well fitted with a dashed straight line with slope
$\delta \simeq 1.74$. (f) Plot of the phase sensitivity exponent
$\delta$ versus $\Delta I_{\rm dc}$ $(=I_{\rm dc}-I^*_{\rm dc})$
for the SN bursting attractor; $I^*_{\rm dc} \simeq 0.2236$.
\label{fig:SNB2}}
\end{figure}

The silent and bursting states for $A_2 = 0.5$ are analyzed in terms of the
largest Lyapunov exponent $\sigma_1$ and the phase sensitivity
exponent $\delta$ in the Poincar\'{e} map. Projections of
attractors onto the $\theta-x$ plane for $I_{\rm dc}=$ 0.21, 0.24,
and 0.29 are given in Figs.~\ref{fig:SNB2}(a)-\ref{fig:SNB2}(c),
respectively. For the silent case, a smooth torus exists in the
$\theta-x$ plane [see Fig.~\ref{fig:SNB2}(a)]. On the other hand,
nonsmooth bursting attractors appear for both SN and chaotic
bursting states, as shown in Figs.~\ref{fig:SNB2}(b) and
\ref{fig:SNB2}(c). A dynamical property of each state is
characterized in terms of the largest Lyapunov exponent $\sigma_1$
(measuring the degree of sensitivity to initial conditions). The
Lyapunov-exponent diagram is given in Fig.~\ref{fig:SNB2}(d). When
passing the bursting transition point $I^*_{\rm dc} (\simeq
0.2236)$, an SN bursting attractor appears. The graph of
$\sigma_1$ for the SN bursting state is shown in black, it is nearly
flat, and its value is negative as in the case of smooth torus.
However, as $I_{\rm dc}$ passes the chaotic transition point $I_{\rm dc}$
$(\simeq 0.2703)$, a chaotic bursting attractor with a positive
$\sigma_1$ appears. Unlike the case of direct transition from a
smooth torus to a chaotic bursting attractor, $\sigma_1$ seems to
increase continuously from zero without jump [compare
Fig.~\ref{fig:SNB2}(d) with Fig.~\ref{fig:SNB1}(c)].
Although SN and chaotic bursting attractors are dynamically different, they both have strange geometry leading to aperiodic complex burstings. To characterize the strangeness of an attractor, we investigate the
sensitivity of the attractor with respect to the phase $\theta$ of
the external quasiperiodic forcing \cite{PD}. This phase
sensitivity may be characterized by differentiating $\bf x$ with
respect to $\theta$ at a discrete time $t=n$. Using
Eq.~(\ref{eq:HR2}), we may obtain the following governing equation
for $\frac {\partial x_i} {\partial \theta}$ $(i=1,2,3)$,
\begin{equation}
  {\frac {d} {dt}} \left( {\frac {\partial x_i} {\partial \theta}}
  \right) =
  \sum_{j=1}^3 {\frac {\partial F_i} {\partial x_j}}
  \cdot{\frac {\partial x_j}   {\partial \theta}} +
  {\frac {\partial F_i} {\partial \theta}},
\label{eq:PSE}
\end{equation}
where $(x_1,x_2,x_3) = (x,y,z)$ and $F_i$'s $(i=1,2,3)$ are given
in Eq.~(\ref{eq:HR2}). Starting from an initial point
$({\bf{x}}(0), \theta(0))$ and an initial value $\partial \bf{x} / \partial
\theta = {\bf 0}$ for $t=0$, we may obtain the derivative values
of $S^{(i)}_n$ $(\equiv
\partial x_i / \partial \theta)$ at all subsequent discrete
time $t=n$ by integrating Eqs.~(\ref{eq:HR2}) and (\ref{eq:PSE}).
One can easily see the boundedness of $S_n^{(i)}$ by looking only
at the maximum
\begin{equation}
\gamma_N^{(i)}({\bf{x}}(0), \theta(0)) = \max_{0 \leq n \leq N}
|S_n^{(i)}({\bf{x}}(0), \theta(0))| \,\,\, (i=1, 2,3). \label{eq:gFtn}
\end{equation}
We note that $\gamma_N^{(i)}({{\bf x}(0), \theta(0)})$ depends on a
particular trajectory. To obtain a ``representative'' quantity
that is independent of a particular trajectory, we consider an
ensemble of randomly chosen initial points $\{ ({\bf x}(0), \theta(0)) \}$, and take the minimum value of $\gamma_N^{(i)}$ with respect to the
initial orbit points \cite{PD}, \begin{equation}
\Gamma_N^{(i)} = \min_{ \{ ({\bf x}(0), \theta(0)) \} } \gamma_N^{(i)}( {\bf
x}(0), \theta(0)) \,\,\, (i=1, 2, 3). \label{eq:GFtn}
\end{equation}
Figure \ref{fig:SNB2}(e) shows a phase sensitivity function
$\Gamma_N^{(1)}$, which is obtained in an ensemble containing 20
random initial orbit points $\{ (x_i(0), y_i(0), z_i(0),
\theta_i(0)); i=1,\dots,20 \}$ which are chosen  with uniform
probability in the range of $x_i(0) \in (-2,2)$, $y_i(0) \in
(-16,0)$, $z_i(0) \in (0,0.4)$, and $\theta_i(0) \in [0,1)$. For
the silent case of $I_{\rm dc}=0.21$, $\Gamma_N^{(1)}$ grows up to
the largest possible value of the derivative $|\partial x_1 /
\partial \theta|$ along a trajectory and remains for all
subsequent time. Thus, $\Gamma_N^{(1)}$ saturates for large $N$
and hence the silent state has no phase sensitivity ({\it i.e.},
it has smooth geometry). On the other hand, for the case of SN
bursting, $\Gamma_N^{(i)}$ grows unboundedly with the same power
$\delta$, independently of $i$,
\begin{equation}
\Gamma^{(i)}_N \sim N^\delta. \label{eq:Gamma}
\end{equation}
Here, the value of $\delta \simeq 1.74$ is a quantitative
characteristic of the phase sensitivity of the SN bursting
attractor for $I_{\rm dc}=0.24$, and $\delta$ is called the phase
sensitivity exponent. For obtaining satisfactory statistics, we
consider 20 ensembles for each $I_{\rm dc}$, each of which
contains 20 randomly chosen initial points and choose the average
value of the 20 phase sensitivity exponents obtained in the 20
ensembles. Figure \ref{fig:SNB2}(f) shows a plot of $\delta$
versus $\Delta I_{\rm dc}$ $(=I_{\rm dc}-I^*_{\rm dc})$. Note that
the value of $\delta$ monotonically increases from zero as $I_{\rm
dc}$ is increased away from the bursting transition point
$I^*_{\rm dc}$ $(\simeq 0.2236)$. As a result of this phase
sensitivity, the SN bursting attractor has strange geometry
leading to aperiodic complex bursting, as in the case of chaotic
bursting attractor.

Using the rational approximation to the quasiperiodic forcing \cite{PD,Kim}, we explain the mechanism for the transition from a silent to an SN
bursting state. For the inverse golden mean, its rational
approximants are given by the ratios of the Fibonacci numbers,
$\omega_k = F_{k-1} / F_k$, where the sequence of $\{ F_k \}$
satisfies $F_{k+1} = F_k + F_{k-1}$ with $F_0 = 0$ and $F_1=1$.
Instead of the quasiperiodically forced system, we study an
infinite sequence of periodically forced systems with rational
driving frequencies $\omega_k$. For each rational approximation of
level $k$, a periodically forced system has a periodic or a
chaotic attractor that depends on the initial phase $\theta_0$ of
the external forcing. Then, the union of all attractors for
different $\theta_0$ gives the $k$th approximation to the
attractor in the quasiperiodically forced system. For this
rational approximation of level $k$, it is sufficient to change
the initial phase $\theta_0$ in the interval $[0,1/F_k)$ in order
to get all possible attracting sets, because the set of all
$\theta$ values fills the whole interval $[0,1)$.

\begin{figure}
\includegraphics[width=\columnwidth]{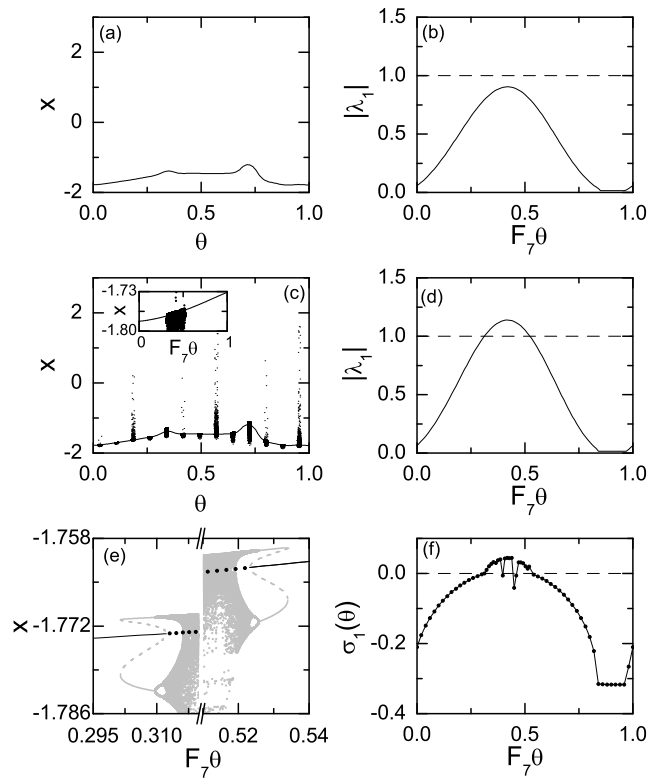}
\caption{Investigation of transition from a smooth torus
(corresponding to a silent state) to an SN bursting attractor
(corresponding to an SN bursting state) in the rational
approximation of level $k=7$ for $A_1=0.5$, $A_2=0.5$, and
$f_1=30$ Hz. In (a) and (c), projections of attractors onto the
$\theta$-$x$ plane are given. (a) Stable smooth torus
corresponding to the silent state for $I_{\rm dc}=0.22$. (b) Plot
of $|\lambda_1|$ vs. $F_7 \theta$ for $I_{\rm dc}=0.22$
[$\lambda_1$: the first stability multiplier (with largest magnitude)
of the $F_7$-periodic orbit]. (c) SN bursting attractor for $I_{\rm dc}=0.222$. A magnified view of the 1st gap is also given in the
inset. (d) Plot of $|\lambda_1|$ vs. $F_7 \theta$ for $I_{\rm
dc}=0.222$. (e) Bifurcation diagram (${\it i.e.},$ plot of $x$ vs.
$F_7 \theta$) for $I_{\rm dc}=0.222$ in the $F_7$-times iterated
Poincar{\'e} map $P^{F_7}$. An $F_7$-periodic orbit (represented
by a black solid line) becomes unstable via a subcritical
period-doubling bifurcation, and then it is denoted by black solid
circles. (f) Lyapunov-exponent diagram for $I_{\rm dc}=0.222$ in
the Poincar{\'e} map. For each randomly chosen initial point, we
get the Lyapunov exponent by following the $10^4$ Poincar\'{e}
maps after the transients of the $10^3$ Poincar\'{e} maps.
\label{fig:RA}}
\end{figure}

We consider the rational approximation of level $k=7$ to the quasiperiodic
forcing of $A_2=0.5$. As shown in Fig.~\ref{fig:RA}(a) for $I_{\rm dc}=0.22$,
the rational approximation to a stable smooth torus (represented
by a black curve), corresponding to a silent state, consists of
stable orbits with period $F_7$ (=13). Figure \ref{fig:RA}(b) shows the magnitude of the first stability multiplier $\lambda_1$ (with the largest magnitude) of the stable $F_7$-periodic orbits in the interval $[0,1/F_7)$. We note that $|\lambda_1|$ varies depending on $\theta$, and all of its values are less than unity. Hence, all $F_7$-periodic orbits for all $\theta$ are stable. However, as $I_{\rm dc}$ passes a threshold value $I_{\rm dc,7}$ $(\simeq 0.2209)$, the smooth torus becomes broken and a nonsmooth bursting attracting set with $F_7$ ``gaps,'' where no stable orbits with period $F_7$ exist, appears. An example is given in Fig.~\ref{fig:RA}(c) for $I_{\rm dc}=0.222$. A magnified view of the 1st gap is given in the inset. For this case, Fig.~\ref{fig:RA}(d) shows $|\lambda_1|$ of the $F_7$-periodic
orbits in the interval $[0,1/F_7)$. In the gap where $0.312\,306<F_7 \theta<0.523\,202$, $|\lambda_1|$ is larger than unity, while in the remaining region of $\theta$, $|\lambda_1|$ is less than unity. Thus, the $F_7$-periodic orbits in the gap become unstable via phase-dependent bifurcations (occurring at specific values of $\theta$), and then chaotic bursting attractors fill the
gap together with regular attractors with periods higher than $F_7$ embedded in small windows. The bifurcation diagram ({\it i.e.}, plot of $x$ vs. $F_7 \theta$ in the $F_7$-times iterated Poincar{\'e} map $P^{F_7}$) for $I_{\rm dc}=0.222$ is given in Fig.~\ref{fig:RA}(e). At both ends of the gap, the
$F_7$-periodic attractor (denoted by a black solid curve) becomes
unstable via a subcritical period-doubling bifurcation when it
absorbs an unstable orbit with doubled period $2 F_7$ (represented
by a gray short-dashed curve). Then, a jump to a chaotic bursting
attractor [developed from the period-doubling cascade of the
stable $2F_7$-periodic orbit (denoted by a gray solid curve)]
occurs. Thus, in Fig.~\ref{fig:RA}(c), the rational approximation
to the whole attractor consists of the union of the periodic
component and the chaotic bursting component, where the latter occupies $F_7$ gaps in $\theta$. Figure \ref{fig:RA}(f) shows the Lyapunov-exponent diagram ({\it i.e.}, plot of $\sigma_1(\theta)$ vs. $F_7 \theta$) for $I_{\rm
dc}=0.222$. (In the gap, chaotic bursting attractors with positive
$\sigma_1$ coexist along with periodic attractors with negative $\sigma_1$ embedded in small windows.) The angle-averaged Lyapunov exponent $\langle\sigma_1\rangle$ [$\langle \cdots \rangle$ denotes the average over the whole $\theta$] is given by the sum of the ``weighted'' Lyapunov exponents of the periodic and chaotic components, $\Lambda_p$ and $\Lambda_c$, (${\it
i.e.,}~\langle\sigma_1\rangle = \Lambda_p + \Lambda_c$), where
$\Lambda_{p(c)} = M_{p(c)} \langle\sigma_1\rangle_{p(c)}$, and
$M_{p(c)}$ and $\langle\sigma_1\rangle_{p(c)}$ are the Lebesgue
measure in $\theta$ and the average Lyapunov exponent of the
periodic (chaotic) component, respectively. Since the periodic
component is dominant, the average Lyapunov exponent
($\langle\sigma_1\rangle \simeq -0.073$) is negative. Hence, the
rational approximation to the whole attractor in
Fig.~\ref{fig:RA}(c) is nonchaotic. We note that Fig.~\ref{fig:RA}(c)
resembles Fig.~\ref{fig:SNB2}(b), although the level $k=7$ is low.
Increasing the level to $k=10$, we confirm that the rational approximations to the whole attractor have $F_k$ gaps (filled with chaotic bursting attractors) which appear via phase-dependent subcritical period-doubling bifurcations and their average Lyapunov exponents are negative. In this way, an SN
bursting attractor appears in the case of quasiperiodic forcing,
as shown in Fig.~\ref{fig:SNB2}(b).

\begin{figure}
\includegraphics[width=\columnwidth]{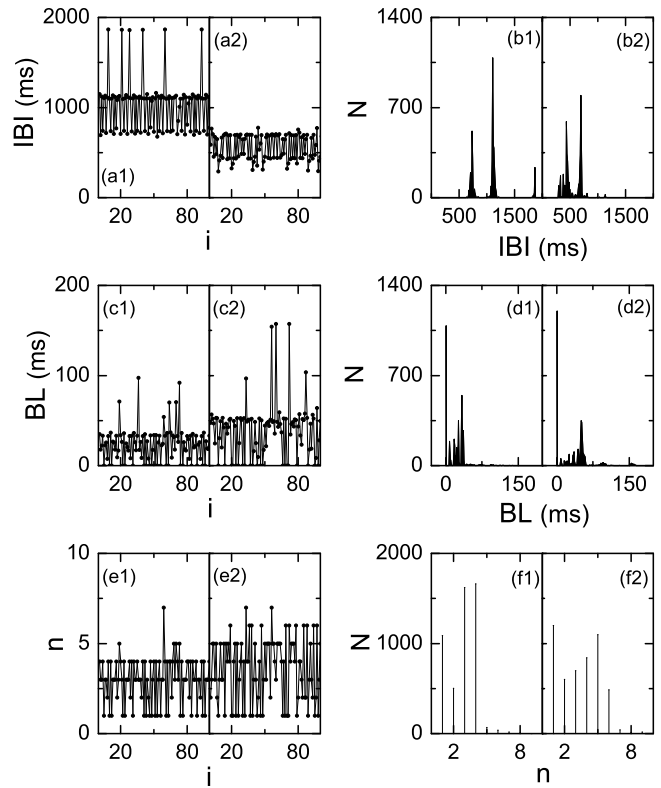}
\caption{Characterization of SN ($I_{\rm dc}=0.24$) and chaotic
($I_{\rm dc}=0.29$) bursting states for $A_1=0.5$, $A_2=0.5$, and
$f_1=30$ Hz. Sequences of interburst intervals (IBIs) for (a1) SN and (a2) chaotic burstings; $i$ represents the bursting index. Histograms of IBIs for (b1) SN and (b2) chaotic burstings. Sequences of bursting length (BL) for (c1)
SN and (c2) chaotic burstings. Histograms of BL for (d1) SN and
(d2) chaotic burstings. Sequences of number of spikes $(n)$ per
burst for (e1) SN and (e2) chaotic burstings. Histograms of $n$
for (f1) SN and (f2) chaotic burstings. For histograms of IBI and
BL, 200 equally-spaced bins are chosen in the range of IBI $\in
(0,2000)$ and BL $\in (0,200)$. We get the number of bursts $(N)$
in each bin among the total number of 5000 bursts (which are also
used for the histogram of $n$). \label{fig:CR}}
\end{figure}

Finally, we characterize the bursting activity [alternating between the silent phase and the active (bursting) phase] in  both cases of SN and chaotic bursting states for $A_2=0.5$. Figures \ref{fig:CR}(a1) and \ref{fig:CR}(a2) show sequences of the interburst intervals [{\it i.e.}, time interval between the first spikes in the neighboring bursts (i.e., active phases)] for the SN and chaotic bursting states when $I_{dc}=$ 0.24 and 0.29, respectively. Both sequences are aperiodic complex ones. Histograms of the interburst intervals for these SN and chaotic bursting states are also given in Figs.~\ref{fig:CR}(b1) and \ref{fig:CR}(b2), respectively. They are multimodal ones. As $I_{\rm dc}$ is increased, heights of peaks for short interburst intervals increase, while those for longer ones decrease. Hence, as $I_{\rm dc}$ is increased, the average interburst interval $\langle \rm{IBI} \rangle$
($\langle \cdots \rangle$ represents the average over a large number of bursts) decreases ({\it i.e.}, the mean bursting rate increases); $\langle \rm{IBI} \rangle \simeq$ 1029 ms and 549 ms for $I_{dc}=$ 0.24 and 0.29, respectively.
The active (bursting) phases are also characterized in terms of the bursting length (BL) ({\it i.e.}, time interval between the first and last spikes in a burst) and the number of spikes (n) per burst. For our cases of the SN and the weakly chaotic burstings, some of the active phases are found to consist of only one spike (i.e., n=1), and hence their BLs are zero. Presence of these active phases with n=1 implies ``weak'' firing activity in such active phases. (This is in contrast to the periodically forced case (of $A_2=0$) where a direct transition from a silent to a chaotic bursting state with all active phases composed of more than one spike occurs.) Sequences of both BL and n are shown in Figs.~\ref{fig:CR}(c1)-\ref{fig:CR}(c2) and Figs.~\ref{fig:CR}(e1)-\ref{fig:CR}(e2). As in the case of the interburst intervals, they are aperiodic complex ones for both the SN and chaotic burstings.  Likewise, their histograms are also multimodal, as shown in Figs.~\ref{fig:CR}(d1)-\ref{fig:CR}(d2) and Figs.~\ref{fig:CR}(f1)-\ref{fig:CR}(f2). With increase in $I_{\rm dc}$, both the average bursting length $\langle \rm{BL} \rangle$ and the average number of spikes $\langle \rm{n} \rangle$ in a burst increase; $\langle \rm{BL} \rangle
\simeq$ 23 ms and 39 ms, and $\langle \rm{n} \rangle \simeq$ 2.9 and 3.4 for
$I_{dc}=$ 0.24 and 0.29, respectively. Thus, both the SN and chaotic bursting states exhibit aperiodic complex burstings, although their dynamics are different (one is chaotic and the other one is nonchaotic). We note that such aperiodic complexity results from the strange geometry of the SN and chaotic bursting states.

\section{Summary}
\label{sec:SUM} We have investigated a dynamical transition from a silent state to a bursting state by varying the dc stimulus $I_{\rm dc}$ in the quasiperiodically forced HR neuron. For this case of quasiperiodic forcing, a transition from a silent state to an SN bursting state (with negative Lyapunov exponent and positive phase sensitivity exponent) has been found to occur when $I_{\rm dc}$ passes a threshold value. With further increase in $I_{\rm dc}$, such an SN bursting state transforms to a chaotic bursting state (with a positive Lyapunov exponent). Thus, a new type of SN bursting states appear between the silent and chaotic bursting states as intermediate ones. This is in contrast to the periodically forced case where a direct transition from a silent state to a chaotic bursting state occurs. Using a rational approximation to the quasiperiodic forcing, the mechanism for the appearance of SN bursting states has been studied. Thus, a smooth torus, corresponding to a silent state, is found to transform to an SN bursting attractor through a phase-dependent subcritical period-doubling bifurcation. Both SN and chaotic bursting states have been characterized in terms of the interburst intervals, the bursting lengths, and the number of spikes per burst. As a result of their strange geometry, both bursting states are found to be aperiodic complex ones, although their dynamics are qualitatively different. Hence, we note that not only chaotic but also SN burstings may become dynamical origin of complex physiological rhythms which are central to life and ubiquitous in organisms.

\begin{acknowledgments}
This research was supported by the Basic Science Research Program through the
National Research Foundation of Korea funded by the Ministry of Education, Science and Technology (2009-0070865).
\end{acknowledgments}


\begin{thebibliography}{}
\bibitem{Chicken} M. R. Guevara, L. Glass, and A. Shrier, Science
{\bf 214}, 1350 (1981); L. Glass, M. R. Guevara, A. Shrier, and R.
Perez, Physica D {\bf 7}, 89 (1983).
\bibitem{Squid} K. Aihara, T. Numajiri, G. Matsumoto, and M.
Kotani, Phys. Lett. A {\bf 116}, 313 (1986); N. Takahashi, Y.
Hanyu, T. Musha, R. Kubo, and G. Matsumoto, Physica D {\bf 43},
318 (1990); D. T. Kaplan, J. R. Clay, T. Manning, L. Glass, M. R.
Guevara, and A. Shrier, Phys. Rev. Lett. {\bf 76}, 4074 (1996).
\bibitem{Squid2} K. Aihara, Scholarpedia 3(5):1786 (2008); see also references therein.
\bibitem{Stoop} R. Stoop, K. Schindler, and L. A. Bunimovich, Neurosci. Res. {\bf 36}, 81 (2000); Nonlinearity {\bf 13}, 1515 (2000).
\bibitem{Glass1} L. Glass and M. C. Mackey, {\it From Clocks to Chaos}
(Princeton University Press, Princeton, 1988).
\bibitem{Glass2} L. Glass, Nature {\bf 410}, 277 (2001).
\bibitem{QO} M. Ding and J. A. S. Kelso, Int. J. Bifurcation Chaos Appl. Sci. Eng. {\bf 4}, 553 (1994).
\bibitem{QKim} W. Lim, S.-Y. Kim, and Y. Kim, Prog. Theor. Phys. {\bf 121}, 671 (2009); W. Lim and S.-Y. Kim, J. Phys. A {\bf 42}, 265103 (2009).
\bibitem{Bursting1} {\it Bursting: The Genesis of Rhythm in the Nervous System} edited by S. Coombes and P. C. Bressloff (World Scientific, Singapore, 2005).
\bibitem{Bursting} E. M. Izhikevich, {\it Dynamical Systems in
Neuroscience} (MIT Press, Cambridge, 2007), p.325.
\bibitem{Greb} C. Grebogi, E. Ott, S. Pelikan, and J. A. Yorke, Physica D
{\bf 13}, 261 (1984).
\bibitem{SNA} U. Feudel, S. Kuznetsov, and A. Pikovsky, {\it
Strange Nonchaotic Attractors} (World Scientific, Singapore,
2006); see also references therein.
\bibitem{PD} A. S. Pikovsky and U. Feudel, Chaos {\bf 5}, 253
(1995).
\bibitem{Kim} S.-Y. Kim, W. Lim, and E. Ott, Phys.\ Rev.\ E
{\bf 67}, 056203 (2003); S.-Y. Kim and W. Lim, J. Phys. A {\bf
37}, 6477 (2004); Phys. Lett. A {\bf 334}, 160 (2005); W. Lim and
S.-Y. Kim, {\it ibid.} {\bf 335}, 383 (2005); {\it ibid.} {\bf
355}, 331 (2006); J.-W. Kim, S.-Y. Kim, B. Hunt, and E. Ott, Phys.
Rev. E {\bf 67}, 036211 (2003).
\bibitem{HR} J. L. Hindmarsh and R. M. Rose, Nature {\bf 296},
162 (1982); Proc. R. Soc. London B {\bf 221}, 87 (1984); {\bf
225}, 161 (1985).
\bibitem{HR1} G. Innocenti and R. Genesio, Chaos {\bf 19}, 023124 (2009);
G. Innocenti, A. Morelli, R. Genesio, and A. Torcini, Chaos {\bf 17}, 043128 (2007).
\bibitem{Lexp} A. J. Lichtenberg and M. A. Lieberman, {\it Regular
and Stochastic Motion} (Springer-Verlag, New York, 1983), p. 283; A. Wolf, J. B. Swift, H. L. Swinney, and J. A. Vastano, Physica D {\bf 16}, 285 (1985).
\end{thebibliography}
\end{document}